\documentclass[12pt]{article}

\usepackage{amsmath,amssymb,url,epsfig,color,cite}
\usepackage{comment}
\definecolor{verdes}{cmyk}{0.92,0,0.59,0.4}
\ifx\pdfoutput\undefined
\usepackage[dvips,bookmarks=false]{hyperref}	
\else
\usepackage{hyperref}	
\fi
\hypersetup{colorlinks,bookmarksopen,bookmarksnumbered,citecolor=verdes,
linkcolor=blus,pdfstartview=FitH,urlcolor=rossos}

\def\m@th{\mathsurround=0pt }
\def\leftrightarrowfill{$\m@th \mathord\leftarrow \mkern-6mu
        \cleaders\hbox{$\mkern-2mu \mathord- \mkern-2mu$}\hfill
        \mkern-6mu \mathord\rightarrow$}

\def\overleftrightarrow#1{\vbox{\ialign{##\crcr
        \leftrightarrowfill\crcr\noalign{\kern-1pt\nointerlineskip}
        $\hfil\displaystyle{#1}\hfil$\crcr}}}

\definecolor{rosso}{cmyk}{0,1,1,0.4}
\definecolor{rossos}{cmyk}{0,1,1,0.55}
\definecolor{rossoc}{cmyk}{0,1,1,0.2}
\definecolor{blu}{cmyk}{1,1,0,0.3}
\definecolor{blus}{cmyk}{1,1,0,0.6}
\definecolor{bluc}{cmyk}{1,1,0,0.1}
\definecolor{verde}{cmyk}{0.92,0,0.59,0.25}
\definecolor{verdec}{cmyk}{0.92,0,0.59,0.15}
\definecolor{verdes}{cmyk}{0.92,0,0.59,0.4}
\definecolor{grigio}{cmyk}{0,0,0,0.07}
\definecolor{rosa}{cmyk}{0,0.1,0.1,0.02}
\definecolor{rosino}{cmyk}{0,0.05,0.05,0.02}
\definecolor{rosas}{cmyk}{0,0.3,0.25,0.05}
\definecolor{celeste}{cmyk}{0.1,0,0,0.02}
\definecolor{giallino}{cmyk}{0,0,0.4,0.02}
\definecolor{rosso}{cmyk}{0,1,1,0.4}
\definecolor{rossos}{cmyk}{0,1,1,0.55}
\definecolor{rossoc}{cmyk}{0,1,1,0.2}
\definecolor{blu}{cmyk}{1,1,0,0.3}
\definecolor{bluc}{cmyk}{1,1,0,0.1}
\definecolor{blucc}{cmyk}{0.7,0.5,0,0}
\definecolor{viola}{cmyk}{0,1,0,0.6}
\definecolor{viola2}{cmyk}{0,1,0.2,0.6}
\definecolor{verde}{cmyk}{0.92,0,0.59,0.25}
\definecolor{verdec}{cmyk}{0.92,0,0.59,0.15}
\definecolor{verdes}{cmyk}{0.92,0,0.59,0.4}
\definecolor{verdino}{cmyk}{0.12,0,0.09,0.05}
\definecolor{giallo}{cmyk}{0,0,1,0}
\definecolor{gialloverde}{cmyk}{0.44,0,0.74,0}

\font\tenrsfs=rsfs10 at 12pt
\font\sevenrsfs=rsfs7
\font\fiversfs=rsfs5
\newfam\rsfsfam
\textfont\rsfsfam=\tenrsfs
\scriptfont\rsfsfam=\sevenrsfs
\scriptscriptfont\rsfsfam=\fiversfs
\def\mathscr#1{{\fam\rsfsfam\relax#1}}

\newcommand{\be}{\begin{equation}}
\newcommand{\ee}{\end{equation}}
\newcommand{\beq}{\begin{equation}}
\newcommand{\eeq}{\end{equation}}

\def\shat{\ifmmode \hat{s}\else $\hat{s}$\fi}
\def\gp2{{g'}^2}
\def\g2{g^2}
\def\g32{g_s^2}

\newcommand{\newc}{\newcommand}

\newc{\gsim}{\lower.7ex\hbox{$\;\stackrel{\textstyle>}{\sim}\;$}}
\newc{\lsim}{\lower.7ex\hbox{$\;\stackrel{\textstyle<}{\sim}\;$}}
\newc{\ie}{{\it i.e.}}
\newc{\etal}{{\it et al.}}
\newc{\mev}{\hbox{\rm\,MeV}}
\newc{\tev}{\hbox{\rm\,TeV}}
\newc{\xpb}{\hbox{\rm\, pb}}
\newc{\xfb}{\hbox{\rm\, fb}}

\newc{\G}{{\cal G}}
\newc{\h}{{\cal H}}
\newc{\D}{{\cal D}}
\newc{\E}{{\cal E}}

\newc{\mtop}{M_t}
\newc{\mbot}{m_b} 
\newc{\mz}{M_Z}
\newc{\mw}{M_W}
\newc{\alphasmz}{\alpha_s(M_Z)}
\newc{\swsq}{\sin^2\theta_W}
\newc{\cwsq}{\cos^2\theta_W}
\newc{\tw}{\tan\theta_W}
\newc{\cw}{\cos\theta_W}
\newc{\sw}{\sin\theta_W}
\newc{\BR}{\hbox{\rm BR}}
\newc{\zbb}{Z\to b\bar}
\newc{\Gb}{\Gamma (Z\to b\bar b)}
\newc{\Gh}{\Gamma (Z\to \hbox{\rm hadrons})}
\newc{\sgn}{\mbox{sgn}}

\newcounter{mysubequation}[equation]

\def\dm2{\delta m^2}
\def\dv2{\delta v^2}

\def\beq{\begin{equation}}
\def\eeq{\end{equation}}
\def\bea{\begin{eqnarray}}
\def\eea{\end{eqnarray}}
\def\slashchar#1{\setbox0=\hbox{$#1$}           
   \dimen0=\wd0                                 
   \setbox1=\hbox{/} \dimen1=\wd1               
   \ifdim\dimen0>\dimen1                        
      \rlap{\hbox to \dimen0{\hfil/\hfil}}      
      #1                                        
   \else                                        
      \rlap{\hbox toc \dimen1{\hfil$#1$\hfil}}   
      /                                         
   \fi}                                         %
\catcode`@=11
\long\def\@caption#1[#2]#3{\par\addcontentsline{\csname
  ext@#1\endcsname}{#1}{\protect\numberline{\csname
  the#1\endcsname}{\ignorespaces #2}}\begingroup
    \small
    \@parboxrestore
    \@makecaption{\csname fnum@#1\endcsname}{\ignorespaces #3}\par
  \endgroup}
\catcode`@=12

\begin{document}

\baselineskip=18pt

\setcounter{footnote}{0}
\setcounter{figure}{0}
\setcounter{table}{0}

\begin{titlepage}
\vspace{.29in}

\begin{center}
{{\bf Top mass determination, Higgs inflation, 
and vacuum stability}}
\vspace{0.5cm}\\
{ Vincenzo Branchina\footnote{branchina@ct.infn.it}, 
Emanuele Messina\footnote{emanuele.messina@ct.infn.it}, 
Alessia Platania\footnote{a.platania90@gmail.com}}
\\
\vspace{0.5cm}
{\em  {Department of Physics and Astronomy, 
University of Catania, and INFN,\\ Via Santa Sofia 64, I-95123 
Catania, Italy}}\\

\vspace{0.5cm}

\end{center}
\vspace{.8cm}

\centerline{\bf Abstract}
\begin{quote}

The possibility that new physics beyond the Standard Model (SM) 
appears only at the Planck scale $M_P$ is often considered. 
However, it is usually argued 
that new physics interactions at $M_P$ do not affect the SM
stability phase diagram, so the latter 
is obtained neglecting these terms. According to this diagram, 
for the current experimental values of the top and Higgs masses,
our universe lives in a metastable state (with very long lifetime), 
near the edge of stability. Contrary to these expectations, however, 
we show that the stability phase diagram strongly depends on 
new physics and that, despite claims to the contrary, 
a more precise determination of the top (as well as of the Higgs) 
mass will not allow to discriminate 
between stability, metastability or criticality of the electroweak 
vacuum. At the same time, we show that the conditions needed for 
the realization of Higgs inflation scenarios (all  
obtained neglecting new physics) are too sensitive 
to the presence of new interactions at $M_P$. Therefore,
Higgs inflation scenarios require very severe fine 
tunings that cast serious doubts on these models. 

\end{quote}

\bigskip
\bigskip

\end{titlepage}

\section{Introduction}

The discovery of the Higgs boson\cite{atlas,cms} is certainly one of 
the most important findings of the last years, the experimental 
data are consistent with Standard Model (SM) predictions, 
and no sign of new physics has been detected. These results
have boosted new interest and work on 
earlier speculations\cite{cabibbo,flores,lindner,bennet,
sherrep,lindsher,arnold,anderson,arnoldvok,ford,sher2,altar,
casas1,espiquiros,casas2,frogniel1,isido,frogniel2} 
that consider the possibility for new phyiscs to show up 
only at very high energies, and for the electroweak vacuum
to be unstable/metastable. A largely explored scenario 
assumes that new physics interactions only appear 
at the Planck scale $M_P$\cite{espigiu,ellisespi,isido,isiuno,isidue,
degrassi}.

To study this scenario, the knowledge of  
the Higgs effective potential $V_{eff}(\phi)$ up to very high 
values of $\phi$ is needed. Due to top loop corrections, 
$V_{eff}(\phi)$ bends down for values of $\phi$ much larger than 
$v$, the location of the electroweak (EW) minimum, and 
develops a new minimum at $\phi_{min} >> v$. 
Depending on Standard Model (SM) parameters, in particular 
on the top and 
Higgs masses, $M_t$ and $M_H$, the second minimum can 
be higher or lower than the EW one. In the first case, the EW 
vacuum is stable. In the second one, it is metastable and
we have to consider the lifetime $\tau$
of the false EW vacuum and compare it with the age of the 
universe $T_U$. If $\tau$ turns out to be larger than $T_U$, 
even though the EW vacuum is not the absolute minimum of 
$V_{eff}(\phi)$, our universe may well be sitting on such 
a metastable (false) vacuum. This is the so called metastability 
scenario. 

\begin{figure}[t]
$$\includegraphics[width=0.7\textwidth]{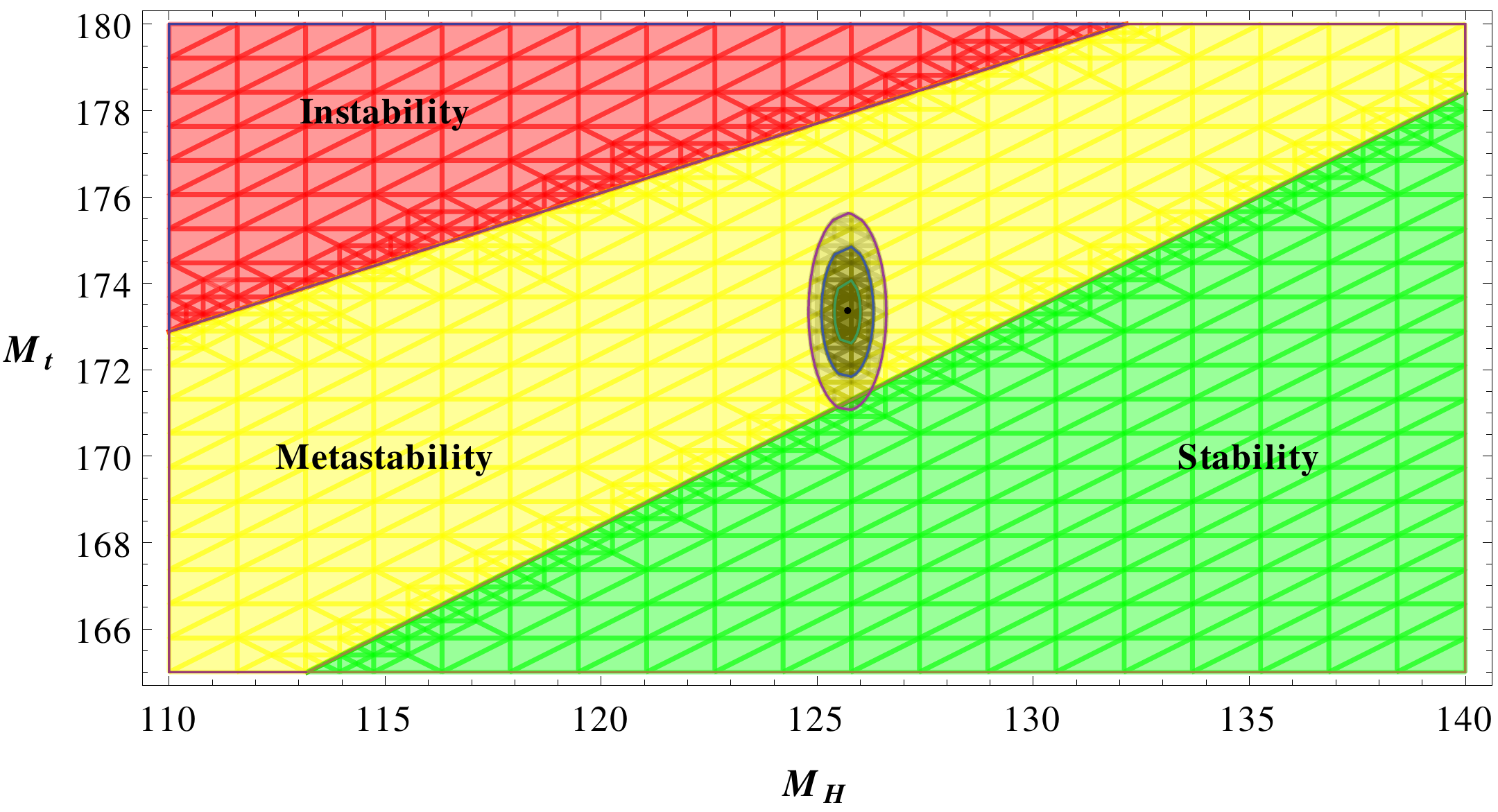}$$
\caption{The stability phase diagram obtained according to 
the standard analysis. 
The $M_H-M_t$ plane is divided in three sectors,  
stability (green), metastability (yellow), and instability 
(red) regions (see text). The dot is for $M_H\sim 125.7$ GeV 
and $M_t\sim 173.34$ GeV (current experimental values). 
The $1\sigma$, $2\sigma$ and $3\sigma$
ellipses are also shown, the experimental uncertainties being
$\Delta M_H=\pm 0.3$ GeV and $\Delta M_t=\pm 0.76$ GeV. 
\label{smphase}}
\end{figure}

This stability analysis is usually presented with
the help of a stability phase diagram in the $M_H-M_t$ 
plane (although, strictly speaking, this is not a phase 
diagram, following the common usage we continue to use 
this expression). The 
standard results\cite{espigiu,ellisespi,isido,isiuno,isidue,degrassi} 
provide the plot shown in fig.\ref{smphase}.
The plane is divided into three different sectors. 
An {\it absolute stability} region (green), where 
$V_{eff}(v) < V_{eff}(\phi_{min})$, a {\it  metastability} 
region (yellow), where $V_{eff}(\phi_{min}) < V_{eff}(v)$, but 
$\tau > T_U $, and an {\it instability} (red) region, where 
$V_{eff}(\phi_{min}) < V_{eff}(v)$ and    
$\tau < T_U $. 
The stability line separates the stability  and 
the metastability sectors, and is obtained for 
$M_H$ and $M_t$ such that $V_{eff}(v) = 
V_{eff}(\phi_{min})$. The instability line
separates the metastability and the instability
regions and is 
obtained for $M_H$ and $M_t$ such that $\tau = T_U $.

According to this analysis, given $M_t \sim 173.34$ GeV 
(central value coming from the 
combination of Tevatron and LHC measurements\cite{accd}) 
and $M_H \sim 125.7$ GeV (average value coming from the 
combination\cite{masistru} of ATLAS and CMS 
results\cite{atla,cm,at2,crn}), the experimental point 
(black dot of fig.\ref{smphase}) lies inside the metastability 
region, 
and, within  $3\,\sigma$,
it could reach and even cross the stability line. 
When it sits on this line, a case named ``critical'', 
the running quartic coupling  $\lambda$ and the beta function 
vanish at $M_P$, $\lambda(M_P) \sim 0$, and 
$\beta (\lambda(M_P)) \sim 0$. From fig.\ref{smphase} we see that 
the black dot is close to the stability line, and the 
``near-criticality'' of the experimental point is considered by 
some as the most important message from the data on the Higgs 
boson\,\cite{degrassi}. The Higgs inflation scenario of\cite{ber},
in particular, strongly 
relies on the realization of the conditions $\lambda(M_P) \sim 0$ 
and $\beta (\lambda(M_P)) \sim 0$. 

Then, given this phase diagram, it is expected that, with the 
help of more refined measurements, we should be able to see 
whether the experimental point sits on the border between 
stability and metastability or is located inside one of these
two regions. More precisely, as the dominant uncertainty comes 
from the value of the top mass,
the expectation is that a more precise determination of $M_t$ 
will finally allow to discriminate between a stable or 
metastable (or critical) EW vacuum\cite{abdel}, \cite{degrassi2}.

The above analysis, however, presents some delicate aspects.
For the central values $M_t \sim 173.34$ GeV and 
$M_H \sim 125.7$ GeV, the Higgs potential $V_{eff}(\phi)$
at $M_P$ is negative (unstable), $V_{eff}(\phi=M_P) <0$, 
and for $\phi > M_P$ it continues to go down for a long while, 
developing the new minimum at $\phi_{min}$ much larger than 
$M_P$, $\phi_{min} \sim 10^{30}\, {\rm GeV}$. Usually, this 
is not viewed as a serious drawback\cite{isido}. In fact, it 
is argued that $V_{eff}(\phi)$ is eventually stabilized 
by new physics interactions present at the Planck scale, 
that should bring the new minimum around $M_P$.
At the same time, it is also argued that the new physics 
terms should not affect the EW vacuum 
lifetime\cite{isido,espigiu}, so the latter is computed 
by considering the unmodified potential $V_{eff}(\phi)$, i.e.
neglecting the presence of new physics.

It has been recently shown, however, that the EW vacuum lifetime 
can be strongly affected by new physics\cite{our}. 
By carring further on this analysis, in the present work we show 
that the expectation that better measurements of the top mass will 
allow to discriminate between a stable, a metastable or a critical
EW vacuum\cite{abdel, degrassi2} is not fulfilled. 
Even very precision measurements of the top mass $M_t$ (as well
as of $M_H$) cannot decide of the EW vacuum stability condition. 
As we will see, the phase diagram of fig.\ref{smphase} can be 
strongly modified by the presence of new physics. 

It is worth to stress, once again, that this phase diagram
is obtained by requiring, on the one hand, that at the Planck 
scale new physics is present and should stabilize the Higgs
potential (otherwise unstable) at this scale, and assuming,
on the other hand, that these new physics interactions have 
no impact in determining the diagram itself. In the light of 
the results of the present work anticipated above, it is 
clear that this phase diagram 
should not any longer be used as the diagram to which 
we refer to decide of the stability condition 
of the EW vacuum. The main actor in determining whether the 
experimental $(M_H$, $M_t)$ point lies in the stability or the 
metastability region is new physics, as 
the latter can strongly affect the stability phase diagram 
(the diagram of fig.\ref{smphase} can be radically changed).    

At the same time, we show that when a specific 
UV completion of the SM, i.e. a specific BSM theory, is 
considered, the stability analysis that takes into account 
the new interactions provides a ``stability test'' for 
the BSM theory under investigation. In this framework, 
precision measurements 
of $M_t$ (and/or $M_H$) provide more stringent constraints 
in the parameter space of the theory.

The rest of the paper is organized as follows. In Section 2
we first review the usual analysis, by considering the case of 
absolute stability of the EW vacuum, and then we  
present the same analysis when new physics 
interactions are taken into account. In Section 3 we do the 
same for the metastability case. In particular, we show how 
the EW vacuum lifetime is computed in the presence of new 
physics. In Section 4, we present the new phase diagrams 
when a specific (toy model) form of new physics is taken into 
account. 
In this section we show how the presence of new physics interactions, 
far from being negligible, can strongly affect the stability 
phase diagram of the EW vacuum. In Section 5 we consider Higgs 
inflation scenarios and apply our analysis to these models, 
showing that the extreme sensitivity to new physics of the conditions 
that need to be realized in order for these models to be viable cast 
serious doubts on them.
Section 6 if for our conclusions.

\section{Stability analysis}

We begin with a short review of the standard stability 
analysis
\cite{isido,espigiu,ellisespi,isiuno,isidue,degrassi}, where new 
physics interactions at the Planck scale are neglected. 
Later, we present the corresponding 
analysis when new physics is taken into account. 

The Higgs potential $V_{eff}(\phi)$ bends down for values of 
$\phi$ larger than $v$ (location of the EW vacuum), and  
develops a new minimum at 
$\phi_{min}$. Depending on $M_H$ and $M_t$, the latter can 
be lower, higher (or at the same height of) the EW minimum.  
Let us normalize $V_{eff}(\phi)$ so that it vanishes at $\phi=v$.
For large values of $\phi$, $V_{eff}(\phi)$ can be written 
as\cite{sher2} 
\be\label{vrgi}
V_{eff}(\phi) \sim \frac{{\lambda_{eff}}(\phi)}{4} \phi^4\, ,
\ee
where $\lambda_{eff}(\phi)$ depends on $\phi$ essentially as 
the running quartic coupling $\lambda(\mu)$ depends on the 
running scale $\mu$. $V_{eff}(\phi)$ is
the renormalization group improved (RGI) Higgs 
potential, and for $\lambda_{eff}(\phi)$ we have the 
corresponding one-loop, two-loops or three-loops expressions.
In the following we consider the up to date 
Next-to-Next-to-Leading-Order (NNLO) 
results\cite{isidue,miha,chety,shapo}.   

For a large range of values of $M_H$ and $M_t$, 
$\lambda_{eff}(\phi)$ has a minimum. Let us call   
${\overline\phi}_{_{M_H, M_t}}$ the point where, for a given 
$(M_H,M_t)$ couple, $\lambda_{eff}(\phi)$ reaches this minimum. 
From Eq.(\ref{vrgi}), we see that the stability line in the 
$(M_H, M_t)$-plane (the line that separates the green and the 
yellow regions in fig.\ref{smphase}) is obtained for those 
couples of values of $M_H$ and $M_t$ such that
\begin{eqnarray}
\lambda_{eff}(\bar\phi_{_{M_H, M_t}})=0\,,
\label{lamco}
\end{eqnarray}
as in this case $V_{eff}(\phi_{min})=V_{eff}(v)$. 
\begin{figure}[t]
$$\includegraphics[width=0.5\textwidth]
{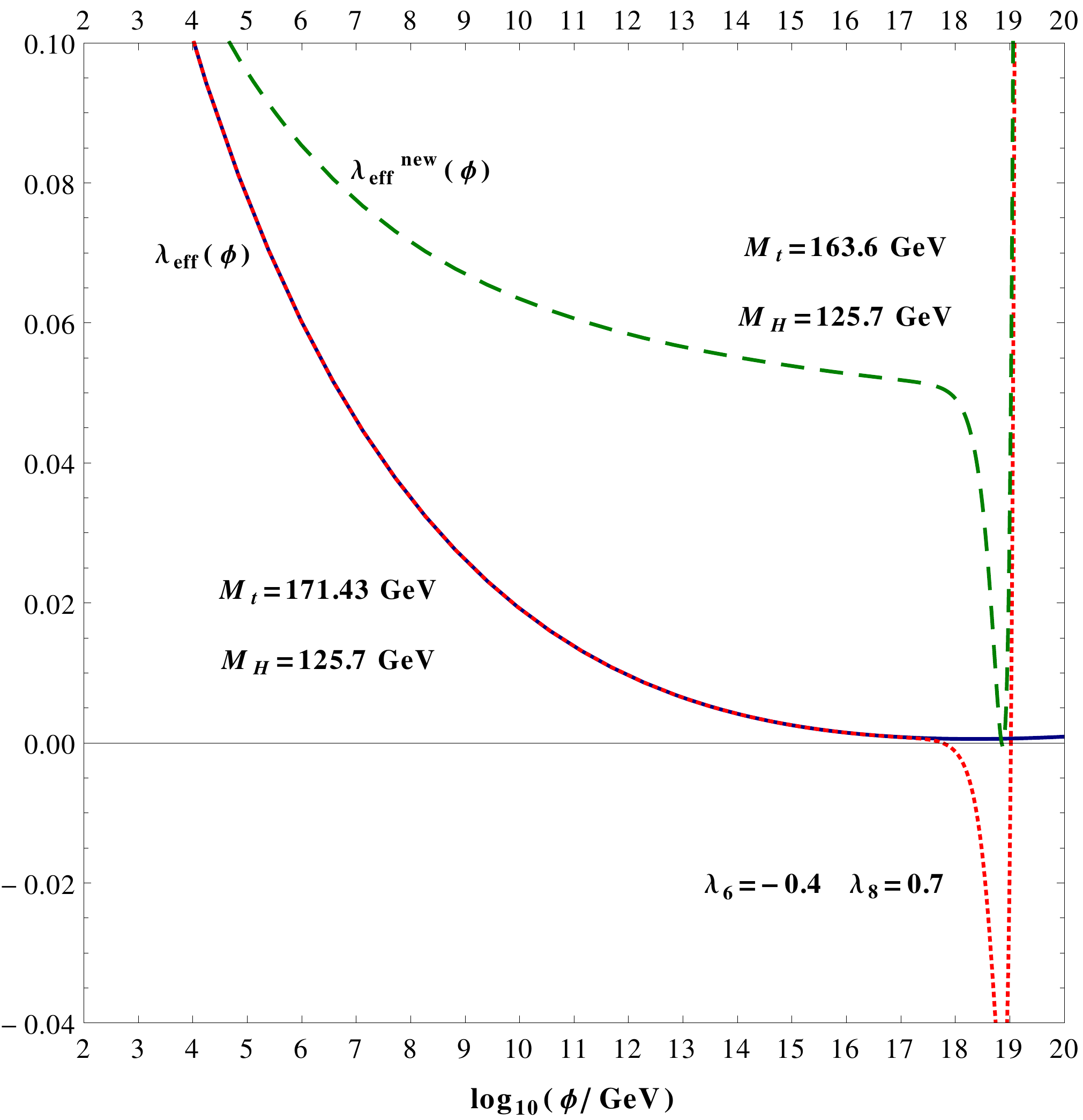}$$
\caption{ The solid (blue) line shows the 
running of $\lambda_{eff}(\phi)$
for $M_H=125.7$\, GeV and $M_t$ 
adjusted so that $\displaystyle\min_\phi\lambda_{eff}(\phi)=0$ 
(see Eq.(\ref{lamco})). 
We get $M_t\sim 171.43$\, GeV, while the minimum is at 
$\bar\phi_{_{M_H, M_t}}\sim 2.22 \cdot 10^{18}$ GeV.
The dotted (red) line shows the running of $\lambda_{eff}^{new}(\phi)$ 
(Eq.\,(\ref{lanew})) for the same values of $M_H$ and $M_t$ 
and for $\lambda_6=-0.4$ and $\lambda_8=0.7$. The minimum 
is formed well below zero.
Keeping fixed the values of $M_H$, $\lambda_6$ and $\lambda_8$, 
the dashed (green) line shows the running of 
$\lambda_{eff}^{new}(\phi)$ for that value of $M_t$ such that 
$\displaystyle\min_\phi\lambda_{eff}^{new}(\phi)=0$.
In this case, we get $M_t= 163.3$ GeV. 
\label{lambda}}
\end{figure}
The solid (blue) line of fig.\,\ref{lambda} provides an example, showing 
the running of $\lambda_{eff}(\phi)$ for $M_H=125.7$ GeV and 
$M_t\sim 171.43$ GeV (see the caption of the figure for an explanation
of these values). The minimum of $\lambda_{eff}(\phi)$
in this case is $\bar\phi_{_{M_H, M_t}}\sim 2.22\cdot 10^{18}$ GeV. 

We are now interested in studying what happens when new physics 
interactions at the Planck scale are taken into account. 
Following \cite{our}, we study the impact of new physics by adding 
to the potential two higher order operators, $\phi^6$ and $\phi^8$. 
With the inclusion of these terms, the classical 
potential $V(\phi)=\frac{\lambda}{4}\phi^4$ becomes
($np$ is for new physics)
\be\label{newpoten}
V_{np}(\phi)=
\frac{\lambda}{4} \phi^4
+\frac{\lambda_6}{6}\frac{\phi^6}{M_P^2}
+\frac{\lambda_8}{8}\frac{\phi^8}{M_P^4}\,,
\ee
with $\lambda_6$ and $\lambda_8$ dimensionless coupling 
constants.

Running the RG equations for all of the SM parameters, including 
$\lambda_6$ and $\lambda_8$, we get the new RGI  potential 
$V_{eff}^{new}(\phi)$,  
\begin{eqnarray}\label{neweffpot}
V_{eff}^{new}(\phi)= V_{eff}(\phi) +
\frac{\lambda_6(\phi)}{6 M_P^2}\xi(\phi)^6\phi^6
+\frac{\lambda_8(\phi)}{8 M_P^4}\xi(\phi)^8\phi^8\,,
\end{eqnarray}
where $\lambda_6(\phi)$ and $\lambda_8(\phi)$, as 
$\lambda_{eff}(\phi)$, are the RG improved couplings, and 
$\xi(\phi)$ comes from the anomalous dimension of $\phi$.
For the purposes of this work, however, it is sufficient to 
keep only the tree level corrections coming from the new 
operators, although the effect of the running can be easily 
taken into account. 

The potential $V_{eff}^{new}(\phi)$ modified 
by the presence of the new physics interactions is then obtained from
Eq.\,(\ref{vrgi}), with $\lambda_{eff}(\phi)$  replaced by
\be \label{lanew}
\lambda_{eff}^{new}(\phi)= \lambda_{eff}(\phi) +
\frac23\,\lambda_6\,\frac{\phi^2}{M_P^2} +
\frac12\,\lambda_8\,\frac{\phi^4}{M_P^4}\,,
\ee
and the stability line is given by  
those values of $\phi$ such that (\ref{lanew}) vanishes.

Let us consider natural (i.e. $O(1)$) values for  
$\lambda_6$ and $\lambda_8$, choosing, for instance, 
$\lambda_6= -0.4$ 
and $\lambda_8=0.7$. The dotted (red) line of fig.\ref{lambda}
is obtained for these values of $\lambda_6$ and $\lambda_8$. 
$M_H$ and $M_t$ are kept fixed to the same values used 
for the solid (blue) line ($\lambda_6=0$, $\lambda_8=0$). 
As expected, for $\phi<<M_P$, the dependence on $\phi$ of 
$\lambda_{eff}^{new}(\phi)$ coincides with that of 
$\lambda_{eff}(\phi)$.
Approaching $M_P$, however, $\lambda_{eff}^{new}(\phi)$ 
becomes negative and develops a minimum well below zero.

If, for comparison, we again consider the case 
$M_H=125.7$ GeV, the corresponding value of $M_t$ that 
brings the SM point (black dot of fig.\ref{smphase}) 
on the stability line
turns out to be sensibly different from the one previously 
determined in the absence of new physics.
The dashed (green) line of fig.\ref{lambda} shows the running of 
$\lambda_{eff}^{new}(\phi)$ for such a value of $M_t$. The 
latter now turns out to be $M_t= 163.60$ GeV and has to be 
compared with the result obtained above in the absence of 
new physics, $M_t= 171.43$ GeV.    


In the present section we set up the tools for 
the determination of the stability line, with and without 
new physics interactions taken into account.  
In Section 4, where the stability phase diagrams are
studied, we will make use of this analysis.  
In the next section, we move to the metastability 
case, i.e. we consider values of $M_H$ and 
$ M_t$ such that the second minimum of $V_{eff}(\phi)$ 
is lower than $V_{eff}(v)$. In these cases, the $EW$ 
minimum is a false vacuum, and we need the tools to 
determine the instability
line, i.e. the boarder between the region where the 
EW vacuum lifetime $\tau$ is larger than the age of the 
universe $T_U$, and the region where $\tau$ is
shorter than $T_U$, the instability region. 

\section{EW vacuum lifetime. Metastability}

The standard analysis of the metastability case is 
performed by computing the 
EW vacuum lifetime $\tau$ with the help of the Higgs 
potential $V_{eff}(\phi)$ that is obtained by considering SM 
interactions only\cite{isido,espigiu,ellisespi,isiuno,isidue}.
As already noted in the Introduction, this is related to 
the expectation that new physics interactions should not affect 
$\tau$.


Referring to \cite{isido,isiuno,isidue,our,mori,inprep} 
for details, we recall here that for a given potential
$V(\phi)$, the general procedure to obtain the tunnelling
time $\tau$ is to look first for the bounce solution (tree level) 
to the euclidean equation of motion\cite{cole1}, and to 
compute then the quantum fluctuations on the top of it\cite{cole2}.
For the Higgs potential $V(\phi)=\lambda\phi^4/4$, once the 
running of the quartic coupling is taken into account, this 
amounts to the following minimization formula 
\begin{eqnarray}\label{tunnelingformula}
\tau= T_U \min_{\mu} \mathcal{\cal T(\mu)} \,
\end{eqnarray}
where $\mathcal{\cal T(\mu)}$ is  
\begin{eqnarray}\label{Tmu}
\mathcal{\cal T(\mu)}&\sim & T_U^{-4} \mu^{-4} 
e^{\frac{8\pi^2}{3|\lambda_{eff}(\mu)|}}\,.
\end{eqnarray}
Together with Eq.(\ref{lamco}), Eq.(\ref{tunnelingformula}) 
is the other key ingredient of the standard stability analysis. 
The phase diagram of fig.\ref{smphase}, in fact, is obtained 
with the help of these equations. The   
stability line (boarder between the stability and the metastability 
regions) in this figure is obtained for those values of
$M_H$ and $M_t$ such that Eq.(\ref{lamco}) is satisfied, 
the instability line (boarder between the metastability region, 
$\tau > T_U$, and the instability region, $\tau < T_U$) 
for those values of $M_H$ and $M_t$ such that $\tau=T_U$.
\begin{figure}[t]
$$\includegraphics[width=0.5\textwidth]
{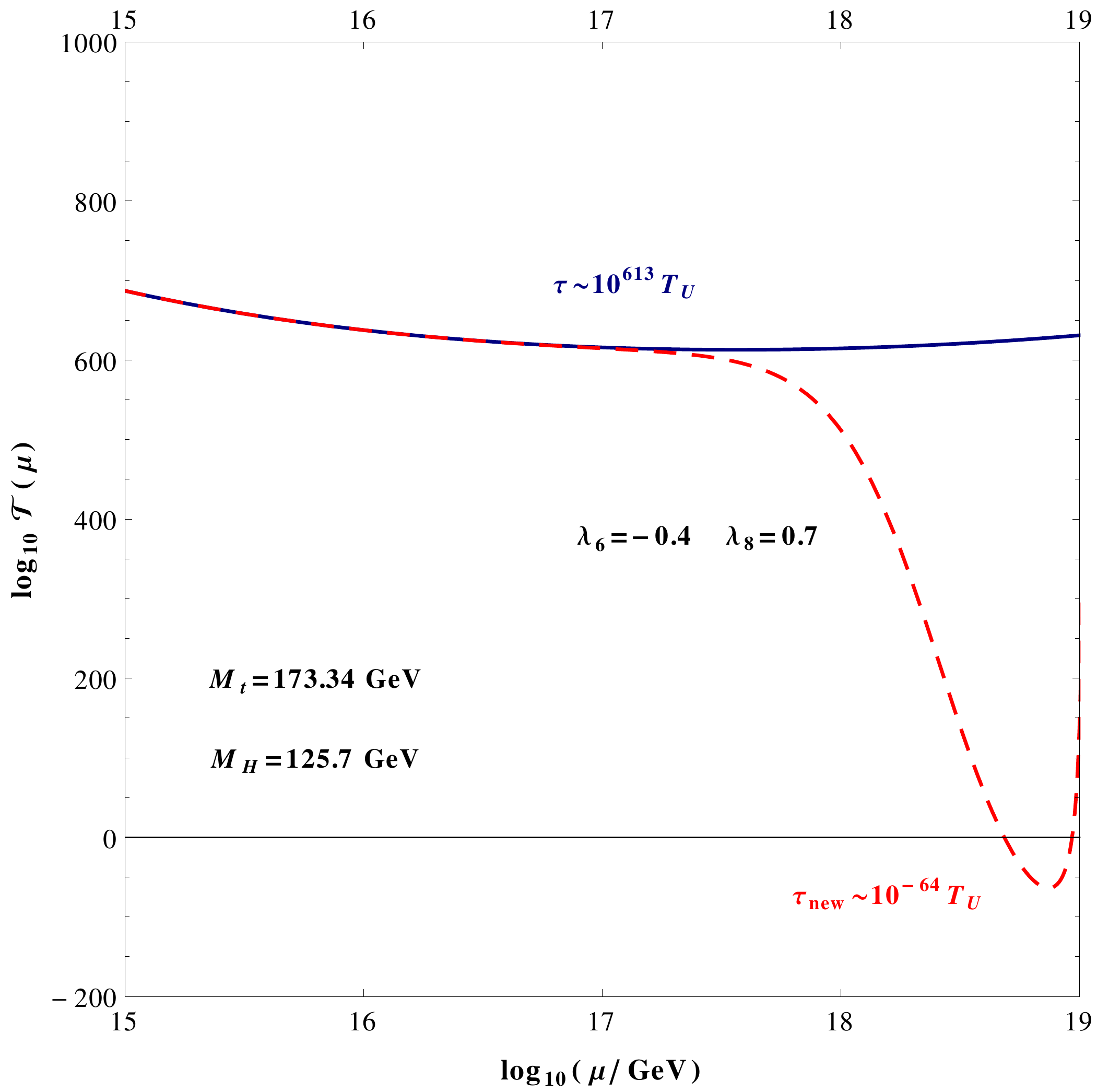}$$
\caption{The solid (blue) line shows the function 
$\log_{10}\cal T(\mu)$ 
for  $M_H=125.7$ \,GeV and $M_t=173.34$\,GeV 
(current experimental values). The minimum forms at 
$\mu \sim 3.6 \cdot 10^{17}$ GeV, that gives $\tau \sim 10^{613} T_U$.
The dashed (red) line shows $\log_{10}{\cal T}_{new}(\mu)$ 
for the same values of $M_H$ and $M_t$ and for $\lambda_6=-0.4$ 
and $\lambda_8=0.7$. In this case the minimum is 
obtained for $\mu\sim 0.62\, M_P$, that gives 
$\tau_{new} \sim 10^{-64} T_U$.  
\label{tunnel}}
\end{figure}

From Eqs. (\ref{tunnelingformula}) and (\ref{Tmu}), the condition  
$\tau=T_U$ is immediately translated into the condition 
\begin{equation}\label{criticalTau}
\min_{\mu}\log_{10}\,{\cal T}(\mu)= 0\,.
\end{equation}
The solid (blue) line of fig.\,\ref{tunnel} is a plot of the 
function $\log_{10}\,\cal T(\mu)$,
for  $M_H=125.7$ GeV and $M_t=173.34$ GeV (central 
experimental values). This function has a minimum at 
$\mu=\mu_{min} \simeq 3.6\cdot 10^{17}$ GeV. The EW vacuum lifetime 
$\tau$ turns out to be $\tau\simeq 10^{\,613}\,T_U$. 

According to this analysis, then, for the central values  
$M_H=125.7$ GeV and $M_t=173.34$ GeV (black dot in fig.\ref{smphase}), 
the EW vacuum is metastable and $\tau$ is much greater than 
$T_U$. 
If we now keep $M_H$ fixed to the value  $M_H=125.7$
GeV and increase $M_t$, we see that  
$\tau$ decreaes and reaches the value $\tau = T_U$ for 
$M_t=178.04$ GeV. This is how the instability 
line is obtained. 

As in the previous section, we want to perform now the 
stability analysis when the presence of new physics is taken into 
account. To this end, we consider again the potential 
$V_{np}(\phi)$ of Eq.(\ref{newpoten}).
We should then begin by considering  
the tree level contribution that comes from the new
bounce solution for the potential (\ref{newpoten}). 
Differently from the previous case ($\phi^4$ term alone), 
due to the presence of the terms $\lambda_6\phi^6$ and $\lambda_8\phi^8$, 
this bounce cannot be found analytically, and we have 
to solve the Euclidean equation of motion
numerically. Then, we should compute the quantum fluctuations 
around the bounce. This complete analysis is presented 
elsewhere\cite{inprep}.




Interestingly, in fact, a good approximate value 
of $\tau$ can be obtained going  
back to Eqs.(\ref{tunnelingformula}) and (\ref{Tmu}) and replacing
in these equations $\lambda_{eff}(\mu)$ with 
$\lambda_{eff}^{new}(\mu)$\cite{our}, where the latter is given in 
Eq.(\ref{lanew}) (in \cite{our} $\lambda_{eff}^{new}$
is called $\lambda_{eff}$). With this replacement, 
the tunnelling time is then given by\cite{flores,arnold}
\begin{eqnarray}\label{newTu}
\tau_{new}= T_U\,{\rm {\min_\mu}} \,{\cal T}_{new}(\mu)\,,
\end{eqnarray}
where ${\cal T}_{new}(\mu)$ is defined as 
${\cal T}_{new}(\mu) = T_U^{-4}\,{{\rm \displaystyle\min_\mu}} \,\left(\mu^{-4}\,
{\rm exp}\,(8\pi^2/(3|\lambda_{eff}^{new}(\mu)|))\right) $\,.

\noindent
We have carefully checked the validity of this approximation against 
the numerical computation of the bounce and of the corresponding
quantum fluctuations\cite{inprep} and found that the two results 
are in good agreement.

The dashed (red) line of fig.\ref{tunnel} is a plot of the function
$\log_{10}{\cal T}_{new}(\mu)$ 
for $\lambda_6=-0.4$, $\lambda_8=0.7$ and for the central 
experimental values of $M_H$ and $M_t$.
We note that, despite of the fact that $\lambda_6$ and 
$\lambda_8$ are natural ($O(1)$ values), and we could 
expect that they would give negligible contribution 
to $\tau$, the impact of these new physics interactions 
on $\tau$ is quite dramatic. The EW vacuum lifetime changes 
from $\tau=10^{613}T_U$ to $\tau_{new}=10^{-64}T_U$.

As for the case of absolute stability considered in the 
previous section, from the example considered above we 
get the strong suggestion that new physics interactions 
at the Planck scale are far from being 
negligible. 
We come back to this point in the next section, where
the phase diagrams are considered.

\section{New physics, new phase diagrams and top mass} 

In the two previous Sections, we set up the tools for  
our analysis. We are now in the position to draw the 
stability phase diagram for different cases, with and 
without new physics interactions taken into account. 

The phase diagram for the case when new physics 
interactions are neglected, as they are supposed to have 
no impact on it, is well known\cite{isidue,degrassi},
and we have reproduced this case in fig.\ref{smphase}. 
From this figure we see 
that, according to this stability analysis, for 
the central values of $M_H$ and $M_t$, the 
EW vacuum is metastable with a lifetime extremely larger 
than the age of the universe ($\tau=10^{\,613}T_U$).
From the same diagram, we also see that, within $3\,\sigma$,
the SM point could reach and even cross the stability line.


Due to the great sensitivity of the results on the stability 
analysis to the value of the top mass, it is usually believed that
a more precise measurement of $M_t$ would provide a definite 
answer to the question of whether we live in the stability region, 
in the metastability region, or at the edge of stability 
(criticality). In particular, it was stressed in\cite{abdel} that
the identification of the measured mass with the pole mass 
is not free of ambiguities (quarks do not appear as asymptotic 
states, and the pole top mass has to be defined with care), 
and that these difficulties can be overcame 
if we refer to the running $M_t^{\overline{MS}}(\mu)$ 
top mass. At the same time, the authors observe that, when 
the translation to the pole mass is appropriately realized,
the error on $M_t$ turns out to be much larger than 
the experimental error usually reported. As a result, 
the Tevatron and LHC results for $M_t$, within two sigma, turn
out to be consistent with stability, metastability, and 
instability at once. 
This analysis seems to point towards the conclusion that 
our knowledge of the stability condition of the EW vacuum  
critically depends on the precise determination of the top 
mass. 

As we will see in a moment, however, while the remarks 
on the top pole mass\cite{abdel} have 
to be seriously taken into account, the expectation that 
a more precise determination of the top mass will 
allow to discriminate between stability or metastability
(or criticality) of the EW vacuum does not seem to be  
fulfilled. Even if new physics interactions show up only 
at the Planck scale, the ``fate'' of our universe (stability 
condition of the EW vacuum) crucially depends on this new physics. 

\begin{figure}[t]
$$\includegraphics[width=0.7\textwidth]
{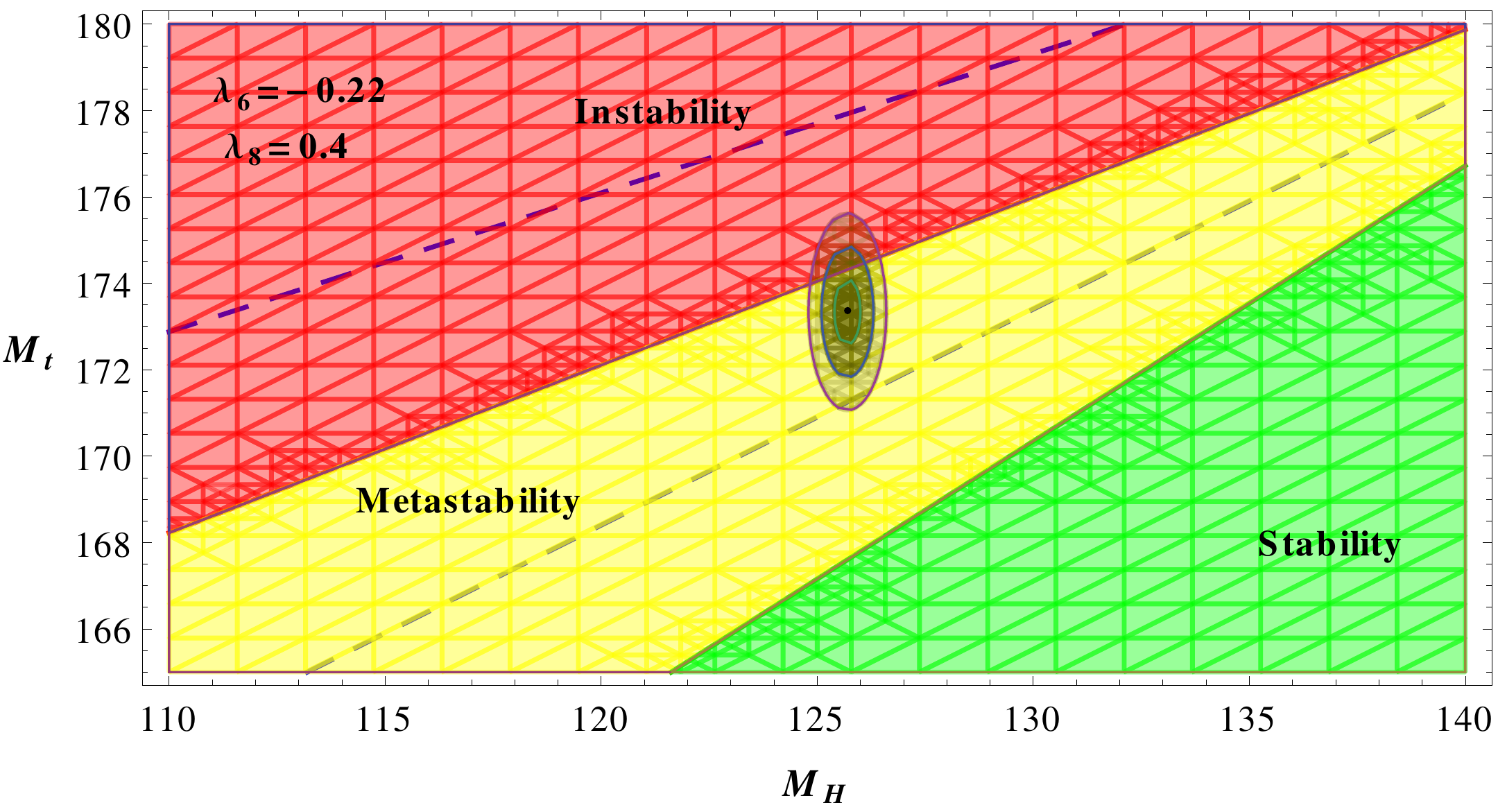}$$
\caption{The stability phase diagram for the potential
$V(\phi)=\lambda\phi^4/4+\lambda_6\phi^6/(6M_P^2)
+\lambda_8\phi^8/(8M_P^4)$ with $\lambda_6=-0.22$ and 
$\lambda_8=0.4$. 
The $M_H-M_t$ plane is divided in three sectors,  
stability, metastability, and instability regions. 
The dot indicates $M_H\sim 125.7$ GeV 
and $M_t\sim 173.34$ GeV. The $1\sigma$, $2\sigma$ and $3\sigma$
ellipses are obtained for the experimental uncertainties 
$\Delta M_H=\pm 0.3$ GeV and $\Delta M_t=\pm 0.76$ GeV. 
The stability and instability lines of fig.\ref{smphase} 
(dashed lines) are reported for comparison.     
\label{smphase2}}
\end{figure}

This is clearly understood if now consider the phase diagram 
that we obtain by following the method presented in the 
two previous sections for a specific choice of new physics
interactions. By taking the potential (\ref{newpoten}) as a 
model of new physics at the Planck scale $M_P$, choosing for 
instance $\lambda_6=-0.22$ and 
$\lambda_8=0.4$, we obtain the phase diagram of 
fig.\ref{smphase2} (the dashed lines are for comparison and 
reproduce the phase diagram of fig.\ref{smphase}). 
As a result of the presence of new physics at the Planck 
scale, the stability and metastability lines move
down. For this choice of $\lambda_6$ 
and $\lambda_8$, the SM point is still in the metastability
region, but its distance from the stability line is larger
than before (more than 5\,$\sigma$).

Comparing the stability phase diagram of fig.\ref{smphase2}
with the one of fig.\ref{smphase}, we clearly see that even 
if the top mass is measured with very high precision, this 
is not going to give any definite indication on the stability
condition of the EW vacuum. As long as we don't know the specific 
form of new physics, we cannot say anything on stability. 
Lowering the error in the determination of $M_t$ is certainly 
important, but definitely not discriminating for 
the stability condition of the EW vacuum. 

\begin{figure}[t]
$$\includegraphics[width=0.7\textwidth]
{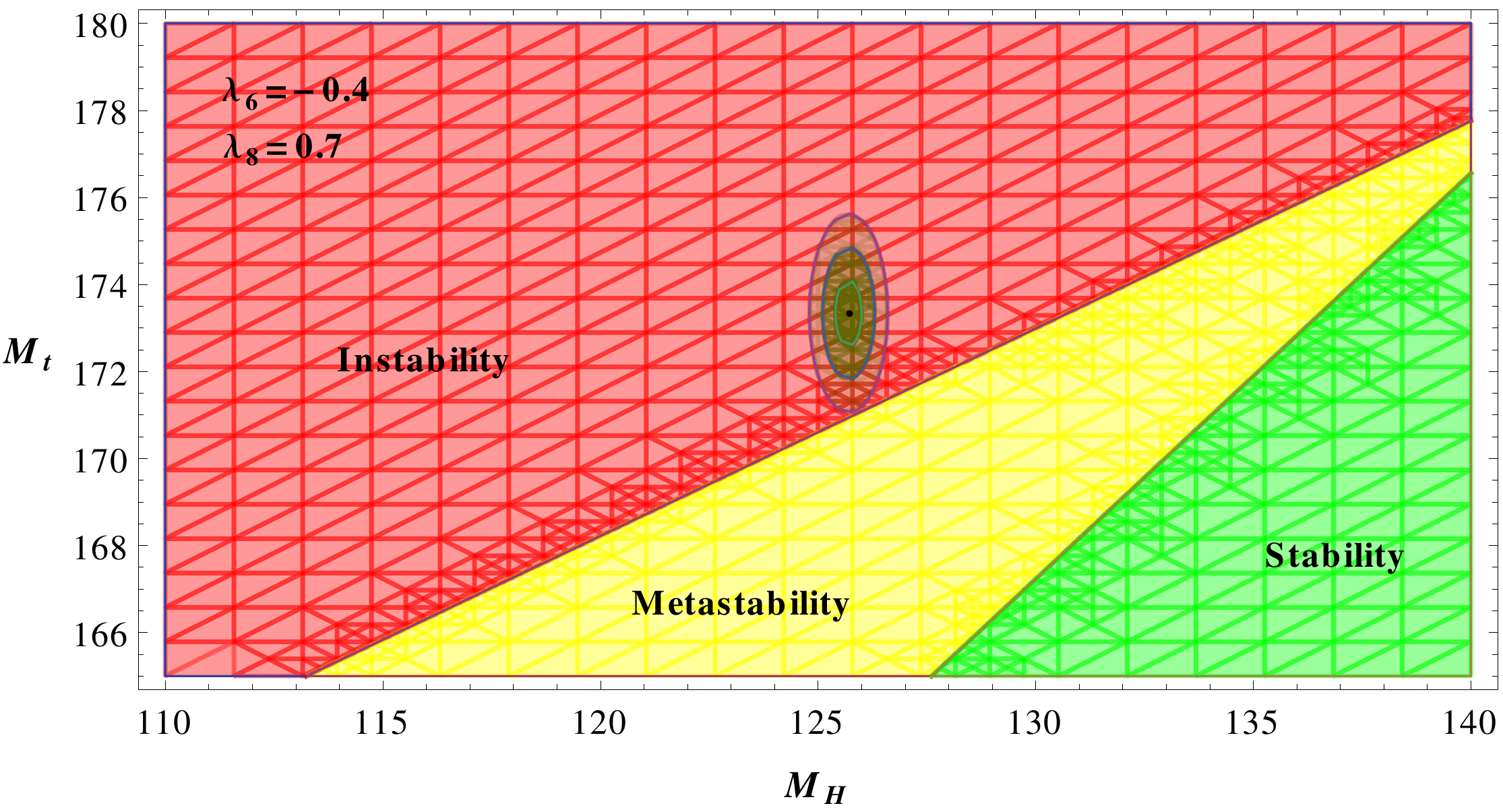}$$
\caption{The stability phase diagram for the potential of 
fig.\ref{smphase2} with $\lambda_6=-0.4$ and $\lambda_8=0.7$. 
The $M_H-M_t$ plane is divided in three sectors,  
stability, metastability, and instability regions. 
The dot indicates  $M_H\sim 125.7$ GeV 
and $M_t\sim 173.34$ GeV. The $1\sigma$, $2\sigma$ and $3\sigma$
ellipses are obtained for the experimental uncertainties 
$\Delta M_H=\pm 0.3$ GeVand $\Delta M_t=\pm 0.76$ GeV.   
\label{smphase3}}
\end{figure}

In order to better appreciate the strong dependence of the 
stability phase diagram on new physics, let us consider  
now a second example, where we use for 
$\lambda_6$ and $\lambda_8$ the values considered in 
Sections 2 and 3, namely $\lambda_6=-0.4$ and $\lambda_8=0.7$. 
It is important to note that (in this as in the previous example) 
we consider natural, i.e. $O(1)$, values for the new 
physics coupling constants. Therefore, the results 
that we get are not driven by an unnatural choice of 
large or small numbers for the (dimensionless) couplings.
They can genuinely come from new physics beyond the SM. 
  
The stability phase diagram for this new case is shown 
in fig.\ref{smphase3}. As compared to the previous case, the 
stability and metastability lines have moved even further 
down, so that, for the central experimental values of $M_H$ 
and $M_t$, the SM point now lies in the instability region. 
\begin{figure}[t]
$$\includegraphics[width=0.45\textwidth]{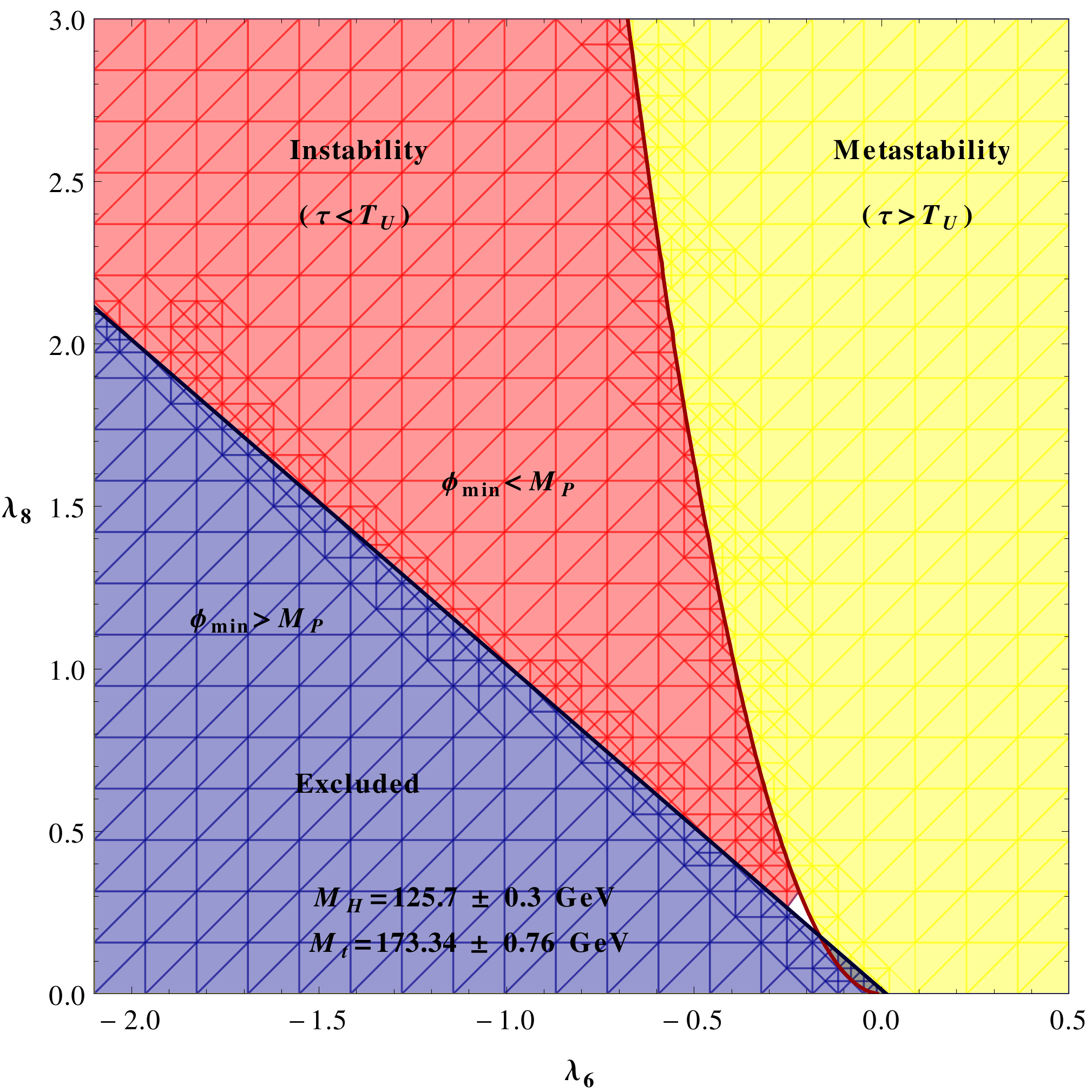}\qquad
\includegraphics[width=0.45\textwidth]{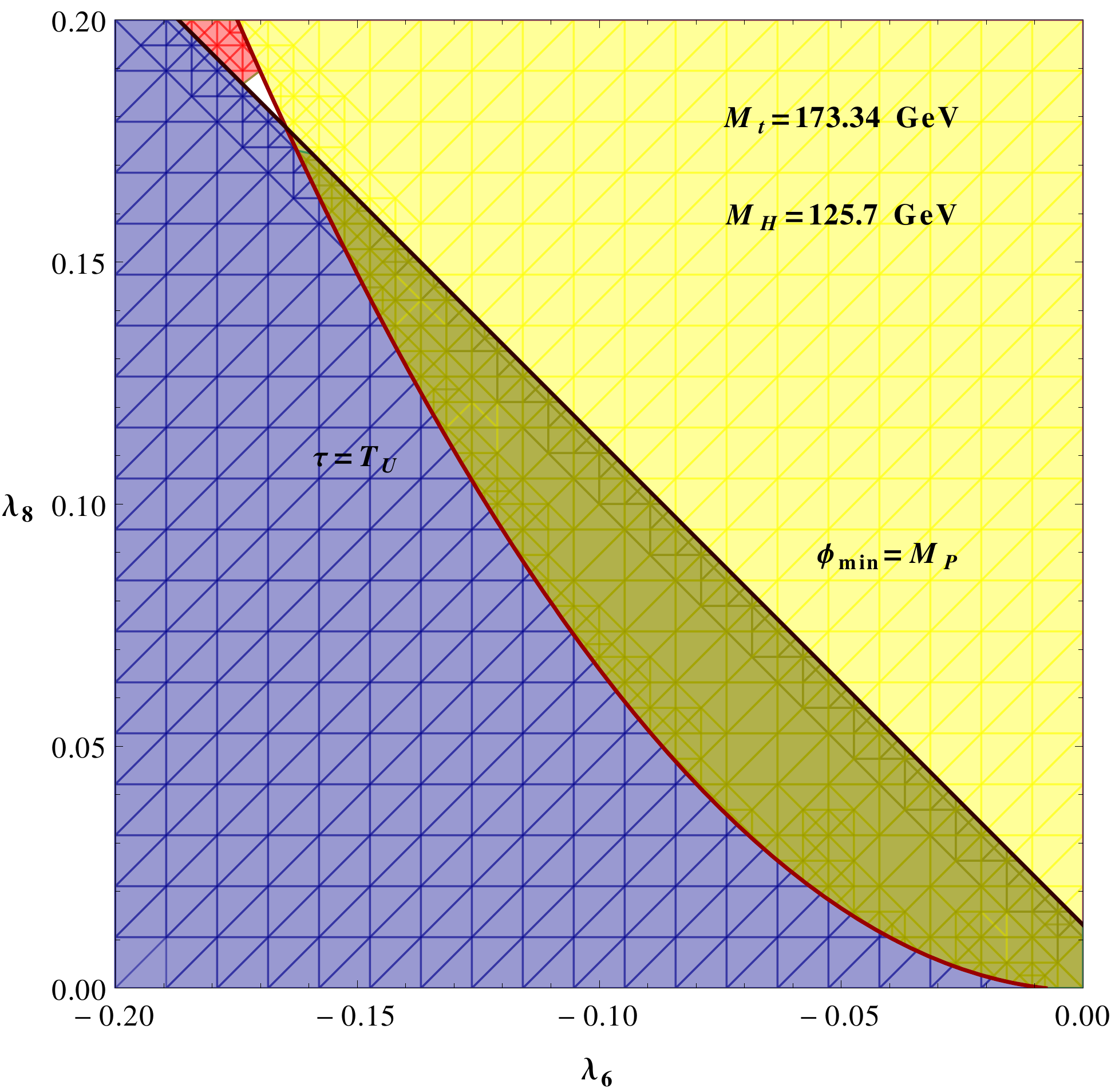}$$
\caption{Left panel - The stability phase diagram in the
$(\lambda_6,\lambda_8)$ plane for the experimental 
values $M_H=125.7$ GeV and $M_t=173.34$ GeV.
The instability (red) and metastability (yellow) regions are 
separated by the $\tau=T_U$ line.
The blue region is obtained for those values of 
$\lambda_6$ and $\lambda_8$ such that $\phi_{min}>M_P$, and the
blue line separates the $\phi_{min}>M_P$ and the $\phi_{min}<M_P$
regions. 
Right panel - Magnification of the region in the lower part of 
the left panel diagram. For those values of 
$\lambda_6$ and $\lambda_8$  
that lie in the dark green region, $\tau$ is 
larger than $T_U$ but the new minimum forms at 
$\phi_{min} > M_P$.
\label{l6l8} }
\end{figure}
\begin{figure}[t]
$$\includegraphics[width=0.5\textwidth]
{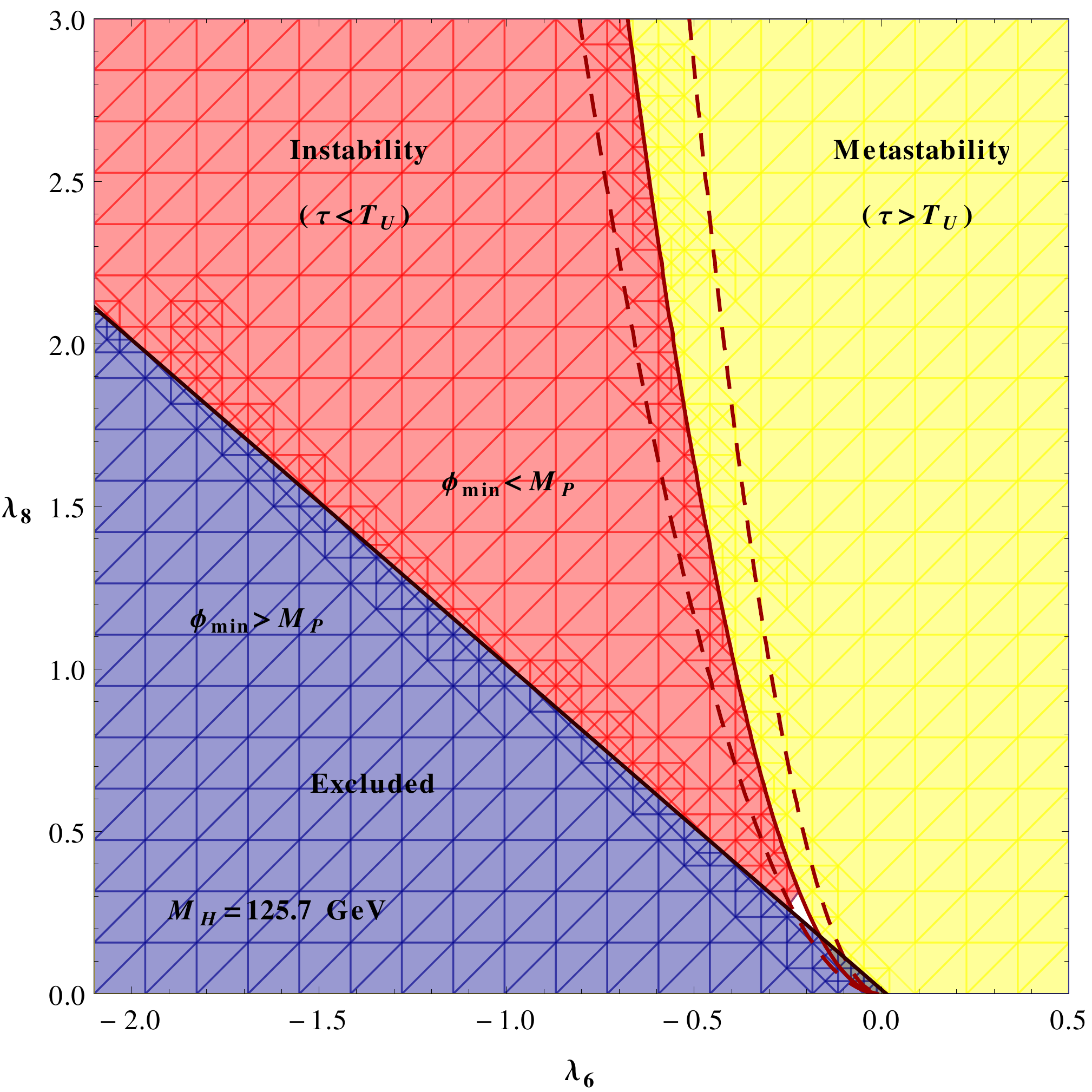}$$
\caption{The same diagram as the left panel of fig.\,\ref{l6l8}, where 
$M_H$ is kept fixed to the value $M_H = 125.7$\,GeV while for $M_t$
a $3\sigma$ excursion from the value  $M_t=173.34$ GeV  
(dashed, red lines) is also considered. 
The experimental uncertainty is $\Delta M_t=0.76$ GeV.  
\label{fascia}}
\end{figure}

Figs. \ref{smphase2} and 
\ref{smphase3} clearly show that the stability phase diagram 
strongly depends on new physics. 
It is then clear that the diagram of fig.\ref{smphase}
(that is the only case considered so far)
is only one out of many different possibilities (see figs. 
\ref{smphase}, \ref{smphase2}, \ref{smphase3}). 
Whether the case of fig.\ref{smphase} , or the case of 
fig.\ref{smphase2}, or another possible case is realized 
(of course the case of fig.\ref{smphase3} is not possible 
for the simple reason that our universe has not decayed!) 
strongly depends on {\it which kind of new physics we have 
at the Planck scale}. 


However, the phase diagram of fig.\ref{smphase} is usually 
presented as if it was the generic result that we obtain 
whenever we assume that the SM is valid all the way up to 
the Planck scale. In particular, referring to this 
phase diagram, it is stated that for the present experimental 
central values of $M_H$ and $M_t$, our universe lives 
within the metastability region, at the edge of the stability 
line\cite{degrassi}, and  that better measurements of $M_H$ 
and $M_t$ will definitely allow to discriminate between 
stability, metastability or criticality for the EW 
vacuum\cite{degrassi2}. 


In the light of what we have shown in the present work, these 
statements appear to be unjustified and misleading. 
What really discriminates between different stability 
conditions for the EW vacuum is {\it New Physics}.
If new physics provides results of the kind 
that we have shown in fig.\ref{smphase2}, the phase diagram 
of fig.\ref{smphase} turns out to be simply wrong, and it 
has definitely nothing to say on the stability condition of 
the EW vacuum. 

Therefore, it is incorrect and misleading to  
refer (as is usually done) to the phase diagram of 
fig.\ref{smphase} as to 
the diagram that provides the picture of the present situation 
for the SM assumed to be valid up to the Planck scale.
We may well have the SM valid up to the Planck scale and, at
the same time, a phase diagram as the one shown in 
fig.\ref{smphase2}. 
The stability phase  
diagram of fig.\ref{smphase}, that is nothing but the well 
known and advertised 
diagram of\cite{isidue,degrassi}, is not universal, it is 
one case out of several different possibilities. 

Let us move now to another important and related lesson that 
we can learn from the above results. 
Going back to the potential 
of Eq.(\ref{newpoten}), let us consider for $M_H$ and 
$M_t$ the current central experimental values,  
$M_H=125.7$ GeV and $M_t=173.34$ GeV, and   
draw the phase diagram of the SM in the 
($\lambda_6$, $\lambda_8$) - plane. The usual 
analysis would tell us that, for these values of $M_H$ and $M_t$,
the EW vacuum is in the metastability region 
(see fig.\ref{smphase}). We have seen, however, that the stability
condition of the EW vacuum depends on new physics,
i.e. $\lambda_6$ and $\lambda_8$ in our present case 
(see also \cite{lalak}).  

The left panel of fig.\ref{l6l8} shows the vacuum 
stability phase diagram in the $(\lambda_6, \lambda_8)$ - plane. 
The line separating the yellow and the red regions is 
the instability line ($\tau=T_U$). The metastability region 
($\tau > T_U$) is on the right side of this 
line, the instability region ($\tau < T_U$) on the left. 
In this diagram we also see the appearence of a new  boarder 
line, the line separating the red and
the blue regions. For couples of values $(\lambda_6, \lambda_8)$ 
in the blue region, the new minimum of the Higgs potential
occurs at $\phi_{min} > M_P$, then we exclude this region of the 
parameter space. In this respect, we note that in the
lower part of the diagram there is a region where 
$\tau>T_U$ (then this region would be allowed from the point of 
view of the vacuum lifetime) but where $\phi_{min}  > M_P$. 
The right panel of fig.\ref{l6l8} shows a zoom on this region 
(dark green area). 

The lesson from the above example is clear. Any beyond SM (BSM) 
candidate theory has to be tested  with the help of
a ``stability test''. A BSM theory is acceptable only if 
it provides either a stable EW vacuum or a metastable one, 
but with lifetime larger than the age of the universe. A phase
diagram of the kind shown in fig.\ref{l6l8} allows 
to determine the regions of the 
parameter space that are permitted by the stability test.

At the same time, it is also clear that a more refined measurement 
of the top (as well as of the Higgs) mass provides more stringent 
constraints on the parameter space. 
Fig.\ref{fascia}, for instance, shows 
the phase diagram in 
the ($\lambda_6$, $\lambda_8$) - plane, where $M_H$ is kept 
fixed to the value $M_H=125.7$ GeV, as for fig.\ref{l6l8}, 
while for $M_t$, in addition to the central value 
$M_t=173.34$ GeV (solid line) of fig.\ref{l6l8}, we also consider 
$3 \sigma$ corrections to $M_t$ (dashed lines), with 
$\Delta M_t=0.76$ GeV \cite{accd}. 
%
%
This is another important lesson of our analysis. 
While not discriminating for the stability issue, a better 
measurement of the top mass has an impact in the determination
of the allowed regions in the parameter space of the theory.


\begin{figure}[t]
$$\includegraphics[width=0.45\textwidth]{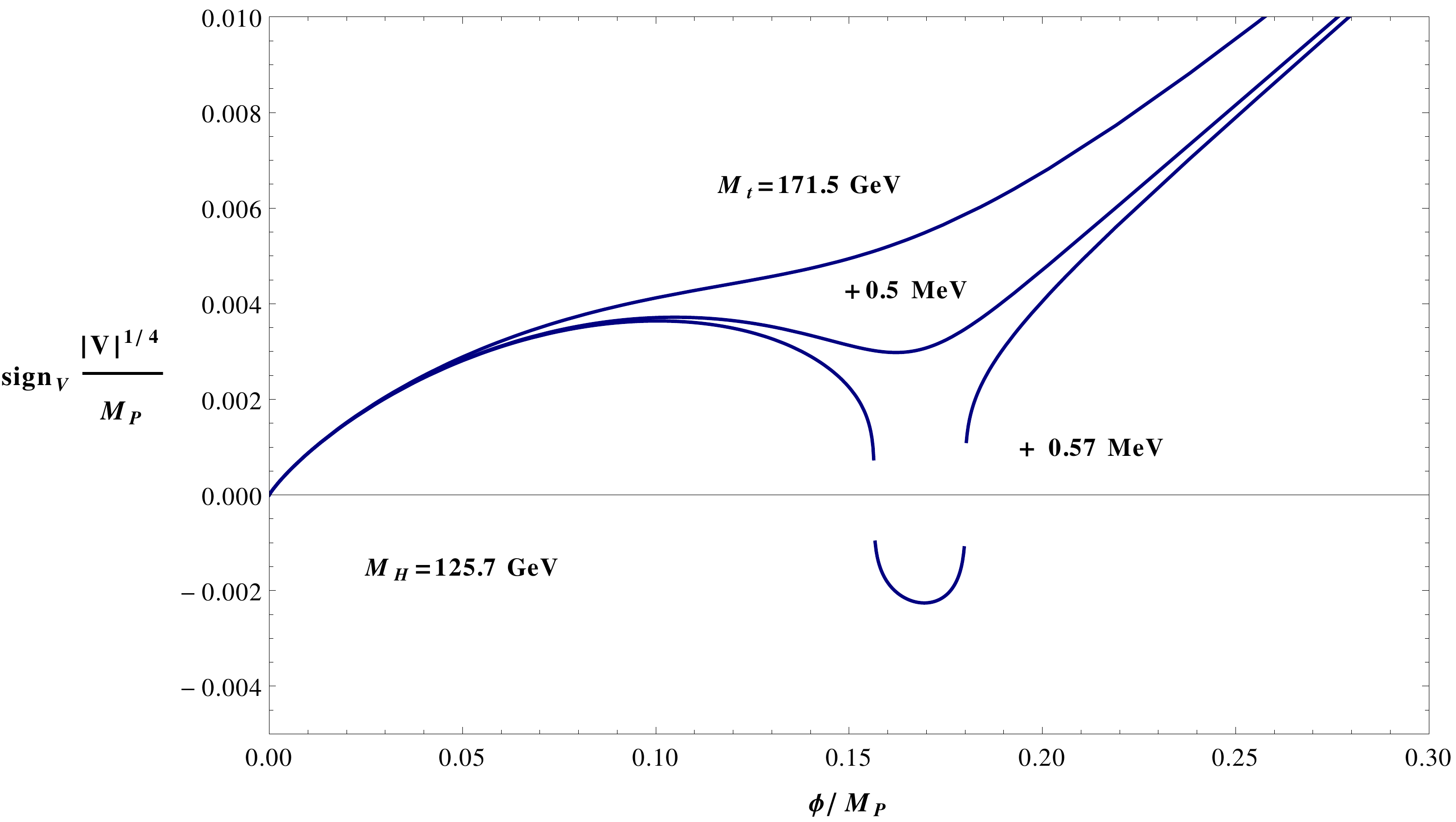}\qquad
\includegraphics[width=0.45\textwidth]{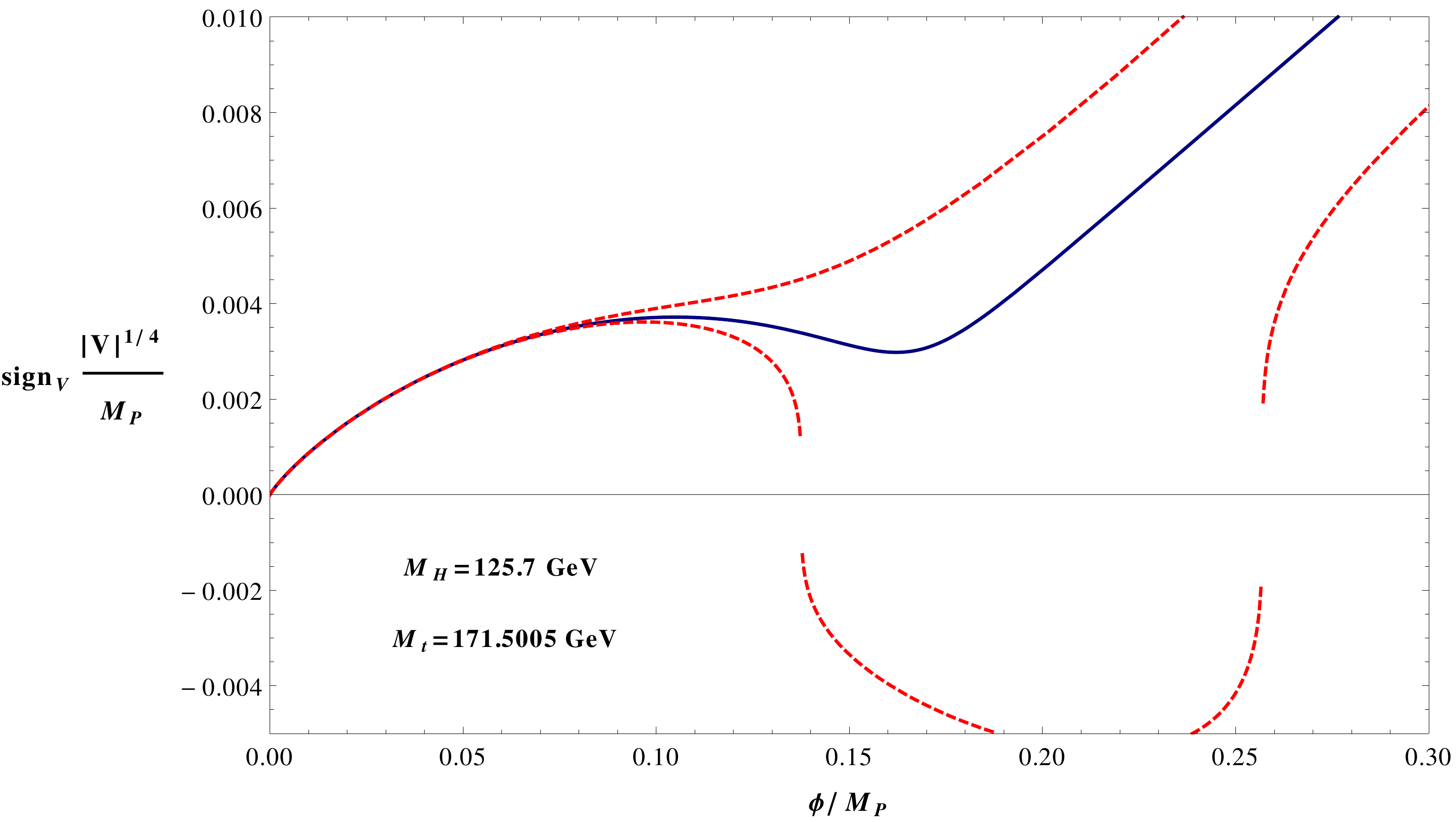}$$
\caption{Left panel - From top to bottom : the $SM$ Higgs 
Potential in $M_P$ units for $M_H=125.7$ GeV and 
$M_t=171.5, 171.5005, 171.50057$ GeV. 
Right panel - The solid (blue) line is the $SM$ Higgs Potential 
for $M_H=125.7 $ GeV and 
$M_t= 171.5005$ GeV. The dashed (red) upper line 
is the Higgs Potential in the presence of 
$\phi^6$ and $\phi^8$, with 
$\lambda_6=2.5\cdot10^{-4}$ and $\lambda_8=1.6\cdot10^{-5}$.
The dashed (red) lower line gives  the same Higgs potential for 
$\lambda_6=-1.17\cdot10^{-4}$ and $\lambda_8=1.6\cdot10^{-5}$. 
\label{infla}}
\end{figure}

\section{Higgs Inflation and new physics} 

Let us come back now to the 
Higgs inflation scenario of\cite{ber}. As noted before, this 
scenario is heavily based on the standard vacuum stability 
analysis. In particular, it requires that new physics shows up 
only at the Planck scale $M_P$, and that the SM lives at
the edge of the stability region, where $\lambda(M_P) \sim 0$ 
and $\beta (\lambda(M_P)) \sim 0$. 
We have seen, however, that new physics interactions,
even if they live at the Planck scale, can strongly change 
the SM phase diagram of fig.\ref{smphase}. The realization of 
the conditions $\lambda(M_P) \sim 0$ and 
$\beta (\lambda(M_P)) \sim 0$ requires such a fine 
tuning\cite{shapber} that even a small grain of new 
physics at the Planck scale can totally destroy the 
picture. We believe that these observation make the 
chance for the realization of the Higgs inflation scenario 
quite low. 

Similar considerations also apply to an alternative 
implementation of Higgs inflation\cite{masinanotari}.  
The idea is that we could have a 
second minimum that is higher than the EW one and such 
that this metastable state could have 
been the source of inflation in the early universe, later 
decaying in the EW stable minimum.
In order to illustrate this 
scenario, together with our comments, we now consider 
the following case. Let $M_H$ have the present 
central value, $M_H=125.7$ GeV, and consider values of 
$M_t$ lower than the central value 
$M_t=173.34$ GeV, actually values of $M_t$ around $M_t\sim 171.5$ GeV.

In the left panel of fig.\ref{infla} we plot the Higgs effective 
potential computed with SM interactions only for 
$M_t=171.5,\,171.5005,\, 171.50057$ GeV.
We see that in the case $M_t=171.5005$ GeV the potential
develops a new (shallow) minimum, higher than the EW one.
As is clear from 
fig.\ref{infla}, there is only a narrow band of values of 
$M_t$ such that this minimum is higher than the 
EW one. In fact (see fig.\ref{infla}), for  
$M_t=171.5$ GeV the minimum disappears, while for 
$M_t=171.50057$ GeV the new minimum is lower that the 
EW one.
 

Needless to say, this proposal has severe intrinsic fine tuning 
problems. Changing the fifth decimal in the top mass, the new 
minimum goes from metastable to stable. But even if we accept 
such a fine tuning for the top mass, and stick on the 
$M_t=171.5005$ GeV value of our example, the presence of even  
a little seed of new physics would be suffient to
screw up the whole picture. This is clearly shown in the right 
panel of fig.\ref{infla}, where the addition of very tiny 
values of new physics coupling constants is able to produce
either the disappearance of the new minimum or the lowering
of this minimum below the EW vacuum.  

\section{Conclusions}

We have shown that the stability condition of the EW vacuum 
and the corresponding stability phase diagram in the 
$(M_H, M_t$) - plane strongly depend on new physics. 
On the contrary, in the past it was thought 
that, given the values of $M_H$ and $M_t$, the stability 
of the EW vacuum could be studied with no 
reference to the specific UV completion of the SM. This 
lead to the believe that the phase diagram of fig.\ref{smphase} 
is universal, that is independent on new physics. As we have 
shown, see figs.\ref{smphase2} and \ref{smphase3}, this is 
not the case. 

This lack of universality has far reaching consequences for 
phenomenology, in particular for model building. As the 
stability condition of the EW vacuum is sensitive to new physics,  
it is clear that any BSM candidate has to pass 
a sort of ``stability test''.
In fact, only a BSM theory that respects the requirement that
the EW vacuum is stable or metastable (but with 
lifetime larger than the age of the universe) can be accepted 
as a viable UV completion of the SM.  

We have also shown that it is incorrect and misleading to refer 
to the phase diagram of fig.\ref{smphase} as if it was the 
snapshot of the present situation for a SM valid up to the 
Planck scale. As we have seen, {\it the SM may well be valid up 
to the Planck scale and still we could have a completely 
different stability phase diagram as compared to the phase 
diagram of fig.\ref{smphase}}, the latter being the only one 
considered in the literature \cite{isidue, degrassi}. 
This phase diagram is not universal,
it depends on the kind of new physics that we have at the 
Planck scale, it is just one case out of different 
possibilities. Therefore, we should no longer refer to this diagram 
as the status of art of our knowledge concerning the stability
condition of the EW vacuum.  

As a consequence of that, it is clear that, despite claims to the 
contrary, with a more precise 
determination of the top mass $M_t$ we will not 
be able to discriminate between stability, metastability, 
or criticality of the EW vacuum. This expectation, 
in fact, is related to the (erroneous) assumption that the 
phase diagram of fig.\ref{smphase} is valid whatever new 
physics we have at the Planck scale. 

At the same time, if we consider a specific UV completion of the 
SM (a specific BSM theory), a more precise knowledge of $M_t$, as 
well as of $M_H$ and of the other parameters, will be important 
to put constraints on the parameter space of the theory. 
In other words, as long as we do not work with a 
specific BSM theory, we cannot draw any conclusions on the stability 
condition of the EW vacuum. Even if new physics shows up 
only at the Planck scale, the ``fate'' of our universe
(that is the stability condition of the EW vacuum) crucially depends 
on the new physics interactions. 

Moreover, it is clear that the same warnings apply to the 
Higgs inflation scenario of \cite{ber}. The latter is heavily 
based 
on the standard analysis, and in particular on the assumptions 
that new physics shows up only at the Planck scale $M_P$ and 
that the EW vacuum is at the edge of the stability region, 
where $\lambda(M_P) \sim 0$ and $\beta (\lambda(M_P)) \sim 0$. 
As we have seen, new physics interactions can strongly change 
the SM phase diagram of fig.\ref{smphase}, thus changing these
relations. In fact, the realization of the conditions 
$\lambda(M_P) \sim 0$ and $\beta (\lambda(M_P)) \sim 0$ 
requires an enormous fine tuning\cite{shapber}, and new 
physics interactions at the Planck scale can easily screw 
up these relations. Other implementations of the Higgs 
inflation idea\cite{masinanotari}, based on the possibility for 
the SM Higgs potential to develop a minimum at lower energies 
(a minimum where inflation could have started in a metastable 
state), are also subject, for the same reasons, to the 
same warnings.

Finally, it is important to note that our analysis on the impact 
of new physics interactions on the stability analysis can be 
repeated even when the new physics scale lies below the Planck 
scale, as could be the case, for instance, of GUT scale.

\end{document}